\documentclass[preprintnumbers,aps,twocolumn,floatfix,superscriptaddress]{revtex4-1}
\usepackage{eso-pic,calc}
\usepackage{graphicx}
\usepackage{amsmath, amsthm, amssymb}
\usepackage{epsfig}
\usepackage{bm}
\usepackage{color}
\usepackage{bbding}
\usepackage{wasysym}

\usepackage[colorlinks=True]{hyperref}  
\hypersetup{
 	colorlinks=true,  
 	linkcolor=blue,  
 	citecolor=blue, 
 	filecolor=magenta,   
 	urlcolor=blue         
 }

\def\etl{$et~al.$~}

\def\beqr{\begin{eqnarray}}
	\def\eqnr{\end{eqnarray}}
\def\beq{\begin{equation}}
	\def\bc{\begin{center}}
		\def\ec{\end{center}}
	\def\eqn{\end{equation}}

\begin{document}

\title{Correlation time in extremal self-organized critical models}
	
\author{Rahul Chhimpa}
\affiliation{Department of Physics, Institute of Science,  Banaras Hindu University, Varanasi 221 005, India}
	
\author{Abha Singh}
\affiliation{Department of Physics, Institute of Science,  Banaras Hindu University, Varanasi 221 005, India}
	
\author{Avinash Chand Yadav\footnote{jnu.avinash@gmail.com}}
\affiliation{Department of Physics, Institute of Science,  Banaras Hindu University, Varanasi 221 005, India}

\begin{abstract}
{We investigate correlation time numerically in extremal self-organized critical models, namely, the Bak-Sneppen evolution and the Robin Hood dynamics. The (fitness) correlation time is the duration required for the extinction or mutation of species over the entire spatial region in the critical state. We apply the methods of finite-size scaling and extreme value theory to understand the statistics of the correlation time. We find power-law system size scaling behaviors for the mean, the variance, the mode, and the peak probability of the correlation time. We obtain data collapse for the correlation time cumulative probability distribution, and the scaling function follows the generalized extreme value density close to the Gumbel function.}
\end{abstract}

\maketitle

\section{Introduction}
At the critical point of a continuous phase transition, such as ferromagnet to paramagnet, the divergence of correlation length leads to the emergence of scaling features in various thermodynamic quantities~\cite{nishimori_2011}. One can reach the critical point by externally tuning the control parameters in the generic critical system. However, the plausible mechanism that can explain scaling features in many natural systems is self-organized criticality (SOC)~\cite{PhysRevLett.59.381, Christensen_2005, Pruessner_2012, marcovic_2014, Watkins2016}, where the systems dynamically evolve to attain a critical state without external fine-tuning. In the critical state, the system responds instantly to a small perturbation through an avalanche of random size. The avalanche sizes lack a characteristic scale and typically follow a power-law distribution with an upper cutoff. The widely cited examples include sandpiles~\cite{PhysRevLett.59.381}, earthquakes~\cite{PhysRevLett.88.178501, PhysRevLett.114.088501}, biological evolutions~\cite{PhysRevLett.71.4083, PhysRevE.53.4723, PhysRevLett.73.906, PhysRevE.108.044109}, neuronal networks~\cite{levina_2007, PhysRevLett.102.118110, Millman_2010}, and rainfall~\cite{ANDRADE1998557}.

While a few SOC systems are solvable~\cite{PhysRevLett.63.1659, dhar_2006, PhysRevE.60.2706}, much of the insight is based on simulation studies of numerous models~\cite{Christensen_2005}. Our concern here is with one specific class of SOC systems that include extremal dynamics, for example, the Bak-Sneppen (BS) model~\cite{PhysRevLett.71.4083} and Robin-Hood (RH) model~\cite{ZAITSEV1992411, Zaitsev_2002}. The BS model describes an ecological evolution of interacting species. In this model, the species, each characterized by a fitness value (a barrier to mutation or extinction), coevolve. Following the Darwinian principle~\cite{darwin1859origin}, the minimum-fit species undergoes a mutation or extinction, along with its connected neighbors (representing food-web interaction). In the original BS model, each species arranged on a ring of $L$ sites is initially assigned a uniformly distributed random number between 0 and 1. The least-fit species and its nearest neighbors are updated by reassigning random numbers from the same uniform distribution. When the system evolves by the simple dynamical rule iteratively, it reaches a critical state exhibiting scaling features in various quantities. The phenomenon of ``punctuated equilibria"~\cite{Gould1993, bak_1997, gupta_2012} also emerges, where species extinction occurs in the form of intermittent bursts separated by a long period of no activity.

The BS model remains extensively studied, uncovering its rich dynamics. To understand the temporal correlation in the model, the fitness signals have been studied, revealing the signature of $1/f^{\alpha}$ noise with the spectral exponent $\alpha \approx 1.2$~\cite{PhysRevE.108.044109, PhysRevE.110.034130}. The scaling feature of the power spectrum sustains up to a lower frequency (inverse of the correlation time), below which the spectrum becomes flat or uncorrelated. The correlation time scales with the system size with a critical exponent $\lambda = 2.51(2)$ for the global fitness (sum of the fitness of all species)~\cite{PhysRevE.108.044109}. The correlation time can also be interpreted differently. It is the time required for the extinction or mutation of all species that existed together and were distributed spatially at some instant in the critical state. Or, when the fitness of each species evolves in the model at least once over all sites, we can term the duration as the fitness \emph{correlation time}. Notice that the temporal memory retained at a particular site no longer survives when the site undergoes an update.

In the BS model, the space-time evolution of least-fit species performs a random walk with a power-law distributed jump size.
Interestingly, the correlation time resembles the cover time of the random walk. In the context of random walks, the time required to cover the whole spatial extent is the full cover time. Chupeau \etl studied the cover time for various types of random walks, ranging from L\'evy to persistent random walks. They revealed that the shifted and scaled cover time follows the Gumbel distribution~\cite{Chupeau2015, barkai2015}.

In other contexts, the search for universal scaling behavior across different percolating systems is an active area of research. In the dynamic perspective of percolation models, the largest cluster size changes as bonds or links are gradually added. Fan \etl examined the largest jump in cluster size as the gap and studied its scaling behavior using extreme value theory (EVT) combined with finite-size scaling (FSS)~\cite{Fan2020}. They found that the scaling functions across different percolation models can follow the Gumbel distribution, suggesting the presence of super-universality. The generalization of the study to higher order gaps for the largest cluster also reveals interesting features~\cite{fang_2024}. However, Feshanjerd and Grassberger~\cite{Mohadeseh_2024} noted that the jumps cannot be considered unbounded and uncorrelated in finite systems.

In the spatial context, the correlation length quantifies the typical distance over which a system exhibits nontrivial correlations. In the thermodynamic limit, the correlation length diverges in the critical Ising model, and the correlation function decays algebraically. Recently, Palmieri and Jensen~\cite{PhysRevResearch.2.013199} investigated critical models, where the instantaneous correlation length, a stochastic variable, fluctuates over time. While the average of this variable provides the correlation length, its distribution offers additional insights into the critical behavior of the system.

\begin{figure}[t]
	\centering
	\scalebox{1}{\includegraphics{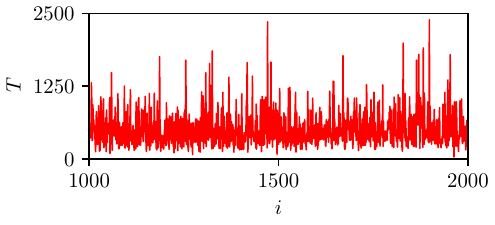}}
	\caption{A typical profile for the correlation time $T$ in the BS model with $L = 2^5$. }
	\label{fig_0}
\end{figure}

We study the correlation time in two extremal SOC models using FSS and EVT. Figure~\ref{fig_0} shows a typical profile of the correlation time for the BS model. Firstly, we examine the system size dependence of various characteristics, such as the mean, the variance, and the mode, that increase in a power law manner. Our scaling arguments reveal that the exponents are the same. Also, the probability of the mode scales with the system size with the same exponent verified numerically. The correlation time probability distribution function shows an unimodal behavior. We also show the data collapse for the cumulative distribution obtained by plotting shifted and scaled correlation time. The scaling function fits reasonably with the generalized extreme value (GEV) distribution with a small positive shape parameter or the Gumbel distribution (cf. Appendix~\ref{app_1}).

The paper's organization is as follows: Section~\ref{sec_2} includes the model definitions. In Sec.~\ref{sec_3}, a scaling argument provides a relation among the scaling exponents for the mean, the variance, and the mode of the correlation time. We also present the results for the distribution function of the correlation time, data collapse, and fitting with EVT. Finally, the results are summarized with a discussion in Sec.~\ref{sec_4}.

\section{Models}{\label{sec_2}}
\subsection{Bak-Sneppen Model}\label{subsec_2_1}
For the simulation of the BS model, we consider a one-dimensional (1D) lattice with periodic boundary conditions. Each site represents a species, assigned a random number from a uniform distribution in the unit interval. This random variable, referred to as the fitness $\eta(x, t)$ or survivability of the species~\cite{PhysRevLett.71.4083}, determines its ability to persist in the ecosystem. The model dynamics consist of two steps: mutation and coevolution. Following the Darwinian principle of ``survival of the fittest'', the species with the lowest fitness is identified and replaced by a new random number drawn from the same distribution (mutation). The coevolution emerges because the extinction or introduction of a new species affects its neighboring species. Specifically, the fitness of the two nearest neighbors of the selected species is also updated through mutation. After sufficient updates, the system reaches a critical state where all species have fitness above a critical value $\eta_c$ [0.667(1) for 1D].

The interaction, or coevolution, is a crucial aspect of the BS model, as the critical state is lost without the interaction, $\eta_c \to 1$~\cite{PhysRevE.80.021132}. A simple change in the interaction rule can yield several variants of the BS model. The least fit species is updated along with only one nearest neighbor in the anisotropic BS (aBS) model~\cite{PhysRevE.58.7141}, one random species in the random-neighbor BS (rBS) model~\cite{PhysRevLett.71.4087}, and one stochastically selected nearest neighbor in the stochastic BS (sBS) model~\cite{PhysRevE.80.021132}. The sBS model belongs to the universality class of the BS model. However, the rBS model corresponds to the mean-field limit.

\subsection{Robin-Hood Model}\label{subsec_2_2}
The Robin-Hood (RH) model, introduced by Zaitsev~\cite{ZAITSEV1992411}, is an extremal model of SOC. It explains the dislocation movement and dry friction models~\cite{PhysRevE.74.066110}. As in the BS model, each site has a random number from the same distribution in the unit interval, called force. The site with the maximum force is considered active, and a stochastic amount of force is removed and distributed to its nearest neighbor sites equally. As the force of the neighbors increases, a correlation in activated sites can be expected. The distance between activated sites shows scale-invariant behavior. Recently, the temporal correlation has been studied in the force fluctuations in the RH model~\cite{Singh_2024}.

\begin{figure}[t]
  \centering
         \scalebox{1}{\includegraphics{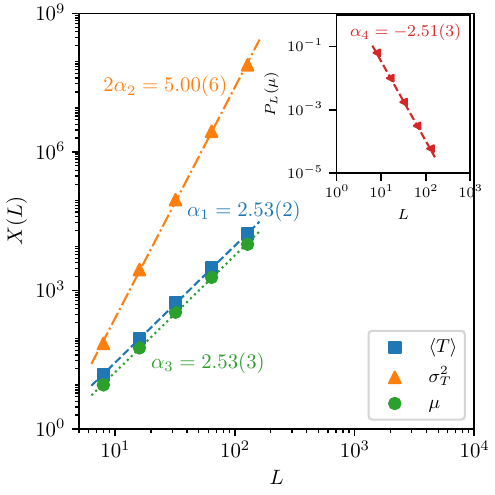}}
  \caption{For the BS model: The system size scaling of the mean $\langle T \rangle \sim L^{\alpha_1}$, the variance $\sigma_{T}^{2} \sim L^{2\alpha_2}$, and the mode $\mu \sim L^{\alpha_3}$ for the correlation time. Inset: The probability for the mode also shows a power scaling $P_L(\mu) \sim L^{-\alpha_4}$. The straight line represents best-fit along with the estimated critical exponents. }
  \label{fig_1}
\end{figure}

\begin{figure}[t]
  \centering
       \scalebox{1}{\includegraphics{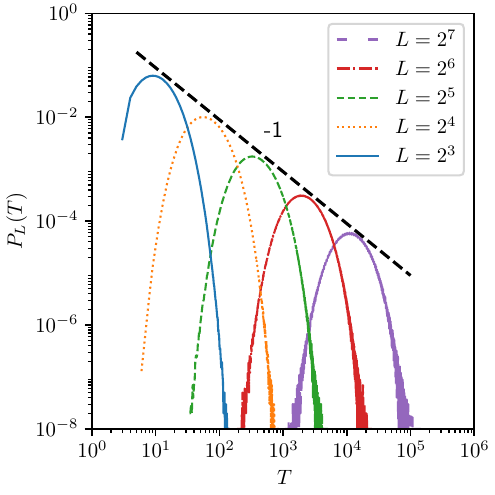}}
  \caption{ The correlation time probability distribution $P_L(T)$ in the BS model for different system sizes $L = 2^3, 2^4, \cdots, 2^7$. Interestingly, the most probable value $P_L(\mu)$ decreases with the system size $L$ in a power law $P_L(\mu) \sim L^{-\alpha_4} \sim T^{-\alpha_4/\alpha_3} \sim T^{-1}$, since $\alpha_4 = \alpha_3$. The dashed straight line provides a numerical confirmation that the scaling exponent is -1.}
  \label{fig_2}
\end{figure}

\section{Results}{\label{sec_3}}
We use the Monte Carlo method to simulate the BS model for generating $M = 10^8$ independent realizations of the correlation time $T$. Notice that the correlation time is a discrete random variable [cf. Fig.~\ref{fig_0}]. To uncover the statistical properties of the correlation time, we examined the system size $L$ dependence. 
As shown in Fig.~\ref{fig_1}, the average and the variance of the correlation time show a system size scaling 
\begin{equation}
\langle T\rangle =  \frac{1}{M}\sum_{i=1}^{M}T_i \sim L^{\alpha_1},
\label{eq_c1}
\end{equation}
and 
\begin{equation}
\sigma_{T}^2 ={\langle T^2\rangle -\langle T\rangle^2} \sim L^{2\alpha_2},
\label{eq_c2}
\end{equation}
where $\langle \cdot \rangle$ denotes ensemble average. The correlation time probability distribution function $P_L(T)$ (cf. Fig.~\ref{fig_2}) shows a unimodal behavior with mode 
\begin{equation}
\mu \sim L^{\alpha_3},
\label{eq_c3}
\end{equation}
and the corresponding probability $P_L(\mu)$ decays with the system size $L$ in a power-law manner as
\begin{equation}
P_L(\mu) \sim L^{-\alpha_4}.
\label{eq_c4}
\end{equation}
The critical exponents discussed above are not independent, and the scaling arguments presented below reveal that all these are basically the same.

To understand the scaling behavior of the distribution function due to finite system size, we introduce a shifted and scaled variable 
\begin{equation}
 u \sim \frac{(T-\mu)}{\sigma_{T}} = \frac{\Delta T}{\sigma_{T}},
 \label{eq_c5}
\end{equation}
 such that the peak of the probability occurs at $u=0$.
In terms of the argument $u$, the expected data collapse or the scaling function is  
\begin{equation}
F(u) =  c \frac{P_L(T)}{P_L(\mu)},
\label{eq_c6}
\end{equation}
where $c$ is a constant.  
Plugging Eqs.~(\ref{eq_c2}), (\ref{eq_c4}), and (\ref{eq_c5}) in Eq.~(\ref{eq_c6}), we can write \begin{equation}
P_L(T) = \frac{1}{c} P_L(\mu) F\left( \frac{T-\mu}{\sigma_{T}}\right) \sim  \frac{1}{L^{\alpha_4}} F\left( \frac{\Delta T}{L^{\alpha_2}}\right). 
\label{eq_c7}
\end{equation}

If the probability $P_L(T)$ [cf. Eq.~(\ref{eq_c7})] is well normalized, then $\int P_L(T)dT = 1$. The normalization condition,
\begin{eqnarray}
\int P_L(T)dT \sim L^{\alpha_2}\int \frac{1}{L^{\alpha_4}}F\left( \frac{\Delta T}{L^{\alpha_2}}\right) d\left(\frac{\Delta T}{L^{\alpha_2}}\right) \nonumber\\ \sim L^{(\alpha_2-\alpha_4)}, \nonumber
\end{eqnarray}
suggests $\alpha_2 = \alpha_4$. 
Similarly, the shifted time average, 
\begin{eqnarray}
\langle \Delta T\rangle \sim  \frac{L^{2\alpha_2}}{L^{\alpha_4}}\int \frac{\Delta T}{L^{\alpha_2}}F\left( \frac{\Delta T}{L^{\alpha_2}}\right) d\left(\frac{\Delta T}{L^{\alpha_2}}\right) \nonumber\\ \sim L^{(2\alpha_2-\alpha_4)} \sim L^{\alpha_1}, \nonumber
\end{eqnarray}
implies a scaling relation $2\alpha_2-\alpha_4 = \alpha_1$ that yields $\alpha_1 = \alpha_2$ because $\alpha_2 = \alpha_4$. Since $u$ is $L$-independent, we get $\langle T\rangle - \mu \sim \sigma_T \sim L^{\alpha_2 = \alpha_1}$, suggesting $\mu \sim L^{\alpha_3 = \alpha_1}$.

\begin{figure}[t]
  \centering
         \scalebox{1}{\includegraphics{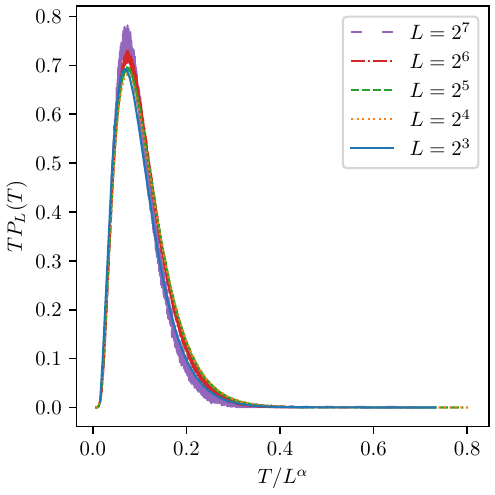}}
  \caption{In the BS model, the data collapse curve for the correlation time probability distribution as $TP_L(T)$ versus $T/L^{\alpha}$ [cf. Eq.~(\ref{eq_c_8})].}
  \label{fig_4}
\end{figure}

Eventually, we find $\alpha_1 = \alpha_2 = \alpha_3 = \alpha_4 = \alpha$. Then, the probability distribution function behaves as
\begin{equation}
P_L(T) \sim \frac{1}{L^{\alpha}} \bar{F}\left( \frac{T}{L^{\alpha}}\right) \sim \frac{1}{T} \bar{G}\left( \frac{ T}{L^{\alpha}}\right).
\label{eq_c_8}
\end{equation}
Clearly, a plot between $T P_L(T)$ and $T/L^{\alpha}$ can provide the scaling function $\bar{G}$ (cf. Fig.~\ref{fig_4}).

\begin{figure}[t]
  \centering
         \scalebox{1}{\includegraphics{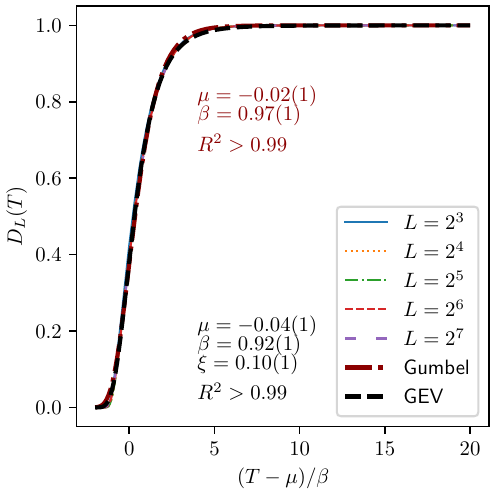}}
  \caption{The data collapse of cumulative probability density corresponds to Fig.~\ref{fig_2} in the BS model. The dark red (thick dashed-dot) represents the fitted curve of the \emph{Gumbel} function with parameters $\mu$ and $\beta$ [cf. Eq.~(\ref{eq_app_1}) with $\xi = 0$]. Here, $\beta \sim \sigma_T$ [cf. Eq.~\ref{eq_app_2})].  To see if there is a significant deviation from the \emph{Gumbel} function, we also fit the scaling function using GEV($\mu, \beta, \xi$) density [cf. Eq.~(\ref{eq_app_1}) with $\xi \ne 0$]. The GEV curve (thick dash) provides a reasonable fit with a small positive shape parameter $\xi$.}
  \label{fig_5}
\end{figure}

\begin{figure}[t]
  \centering
         \scalebox{1}{\includegraphics{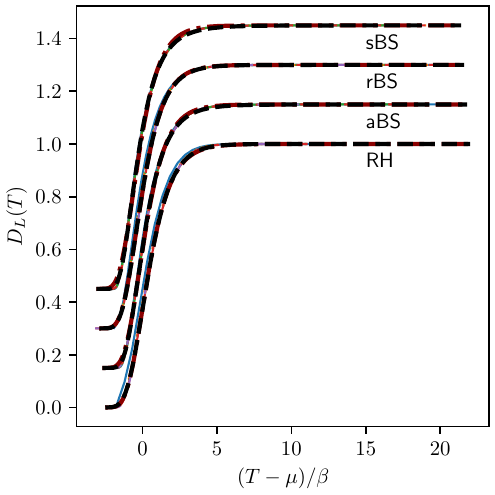}}
  \caption{Same as Fig.~\ref{fig_5}, but for different variants of the BS model. (cf. Table~\ref{tab2} for the fitting parameters). To avoid overlaps, we shift the curves vertically and horizontally by constants.}
  \label{fig_6}
\end{figure}

To smoothen the statistics, we consider the cumulative probability function $D_L(T) = \int_{-\infty}^{T}P_L(T)dT$. As shown in Fig.~\ref{fig_5}, we find a reasonably good data collapse for $D_L(T)$ in the BS model. We fit the scaling function with the Gumbel and GEV density functions [cf. Eq.~(\ref{eq_app_1})] following the method discussed in Appendix~\ref{app_2}. We also examine the correlation time statistics in several variants of the BS model and find qualitatively similar scaling features (cf. Fig.~\ref{fig_6}). We present a summary of the critical exponents in Table~\ref{tab1} and the fitting parameters of the scaling function in Table~\ref{tab2}. As in Table~\ref{tab1}, the estimated exponents $\alpha$ agree well with the correlation exponents $\lambda$, computed in previous studies following a different approach. The fitting parameters shown in Table~\ref{tab2} for GEV indicate a small positive shape parameter consistent with the Gumbel function.

\begin{table}[]
\centering
\begin{tabular}{|c|cccc|c|}
\hline
~~Model~~&$\alpha_1$&$\alpha_2$&$\alpha_3$&$\alpha_4$&$\lambda$\\
\hline
BS&2.53(2)&2.50(3)&2.53(3)&2.51(3)&2.51(2)\\
aBS&1.82(2)&1.78(1)&1.83(3)&1.80(1)&1.80(2)\\
rBS&1.28(2)&1.06(2)&1.35(4)&1.09(2)&1.04(2)\\
sBS&2.30(8)&2.2(1)&2.34(7)&2.24(9)&2.37(7)\\
%eBS&2.21(4)&2.20(3)&2.24(4)&2.21(5)&2.13(3)\\
RH&2.04(1)&2.12(1)&1.98(4)&2.09(1)&2.2(1)\\
%\hline
%BS&2&3.36(6)&3.20(5)&3.42(6)&3.20(6)&3.04(3)\\
\hline
\end{tabular}
\caption{For different variants of the BS model, the critical exponents describe various statistical properties of the correlation time [cf. Eqs.~(\ref{eq_c1})-(\ref{eq_c4})]. For a comparison, we tabulate the correlation exponent $\lambda$ as reported in previous studies~\cite{PhysRevE.108.044109, Singh_2024} obtained through a different approach. }
\label{tab1}
\end{table}

\begin{table}[]
\centering
\begin{tabular}{|c|cc|ccc|}
\hline

&\multicolumn{2}{c|}{Gumbel}&\multicolumn{3}{c|}{GEV}\\
~~Model~~&$\mu$&$\beta$&$\mu$&$\beta$&$\xi$\\
\hline

BS&-0.02(1)&0.97(1)&-0.04(1)&0.92(1)&0.10(1)\\
aBS&-0.02(1)&0.99(1)&-0.03(1)&0.95(1)&0.07(1)\\
rBS&-0.02(1)&0.99(1)&-0.02(1)&0.98(1)&0.02(1)\\
sBS&-0.02(1)&0.95(1)&-0.04(1)&0.89(1)&0.11(1)\\
%eBS&-0.02(1)&0.99(1)&-0.03(1)&0.95(1)&0.06(1)\\
RH&-0.02(1)&1.05(1)&-0.02(1)&1.06(1)&0.02(1)\\
%\hline
%BS&2&-0.03(1)&1.02(1)&$>0.99$&-0.01(1)&1.02(1)&0.01(1)&$>0.99$\\

\hline

\end{tabular}
\caption{For different variants of the BS model, the fitting parameters (cf. Appendix~\ref{app_2}) that describe the scaling function of the correlation time cumulative probability function. In all cases, the goodness of fit is $R^2>0.99$.}
\label{tab2}
\end{table}

\section{Summary}{\label{sec_4}}
In summary, we have studied the Bak-Sneppen model (along with its variants) and the Robin Hood model, displaying self-organized criticality. In the models, we examined the fitness correlation time, the duration required to update all species at least once in the entire space. We use the Monte Carlo simulation method to generate a set of realizations with different system sizes and compute various statistical characteristics of the correlation time, such as the mean, the variance, and the mode. These characteristics follow a system size scaling with the same critical exponent. The correlation time probability distribution function shows an unimodal behavior, with the peak probability decaying in a power law manner. Employing finite-size scaling combined with extreme value theory, we obtain the scaling function for the cumulative probability distribution function of the correlation time. The data collapse fits well with the GEV function with a positive small shape parameter or the Gumbel function.

The correlation exponent $\alpha$ estimated in the present work for different variants of the BS model consistently agrees with the previously reported values of the exponent $\lambda$~\cite{PhysRevE.108.044109, Singh_2024} following a different route. A signature of universal behavior arises as the data collapses follow the Gumbel function across different variants in the BS model. It would be interesting to apply the presented methodology to other critical systems to understand the scaling properties of the correlation time.  

The Hamming distance, or damage spread, measures the sensitivity to initial conditions. For the BS model, it increases in a power-law manner, indicating weak sensitivity to initial conditions~\cite{Tamarit1998}. For a fixed system size, different configurations exhibit weak correlations. Although a marginal deviation from the Gumbel distribution is expected, it seems undetectable.

\section*{ACKNOWLEDGMENTS}
RC acknowledges financial support through the Junior Research Fellowship, UGC, India. AS is supported by Banaras Hindu University through a fellowship (Grant No. R/Dev./Sch/UGC Non-Net Fello./2022-23/53315), and ACY acknowledges a seed grant under the IoE (Seed Grant-II/2022-23/48729). 

\appendix
\section{GEV distribution}\label{app_1}
EVT offers a systematic study of extreme events, which are rare but can have devastating consequences~\cite{Hansen_2020}. Therefore, the extreme events are of practical concern. EVT focuses on understanding extreme deviation from its mean value~\cite{MAJUMDAR20201}. Typical distributions are the Generalized Extreme Value and the Generalized Pareto. The GEV follows the maximum of a set of random variables, and it combines three different functions, i.e., Gumbel, Fr\'echet, and Weibull, based on a shape parameter $\xi$. When $\xi$ tends to 0, it reduces to the Gumbel function. For example, the statistical measurements of extreme weather events reveal that the distributions of the magnitudes of the extreme events satisfy the Gumbel distribution~\cite{Yao_2022}. The generalized Pareto distribution explains extreme values above a threshold value.

Given the location parameter $\mu$, the scale factor $\beta$, and the shape parameter $\xi$, the generalized extreme value probability density GEV($\mu, \beta, \xi$) of extreme random variables is 
\begin{equation}
p(z) = \frac{1}{\beta}t(z)^{\xi+1}\exp{\left[-t(z) \right]},\nonumber
\end{equation}
where 
\begin{equation}
t(z) = \begin{cases} \left[1+\xi \left( \frac{z-\mu}{\beta}\right) \right]^{-\frac{1}{\xi}}, ~~~~\xi\neq 0, \\ \exp\left(- \frac{z-\mu}{\beta} \right), ~~~~~~~~~~ \xi = 0, 
\end{cases}\nonumber
\end{equation}
and the cumulative density function is
\begin{equation}
G(z) = \begin{cases} \exp\left\{-\left[1+\xi\left( \frac{z-\mu}{\beta}\right) \right]^{-\frac{1}{\xi}} \right\}, ~~~~\xi\neq 0, \\ \exp\left\{ -\exp\left(- \frac{z-\mu}{\beta} \right)\right\}, ~~~~~~~~~~~ \xi = 0. \end{cases}
\label{eq_app_1}
\end{equation}

The GEV($\mu, \beta, \xi = 0$) corresponds to the Gumbel distribution for which the mean is $ \langle z \rangle = \mu  + \beta \gamma$, where $\gamma  \approx 0.57721$ is the Euler-Mascheroni constant and the variance is
\begin{equation}
\sigma^2 = \frac{1}{6}\pi^2\beta^2.
\label{eq_app_2}
\end{equation}

\section{Curve fitting method}\label{app_2}
For the GEV and Gumbel curve fitting in Fig.~\ref{fig_5}, we employed a non-linear least squares fitting method, the Levenberg–Marquardt algorithm (LMA)~\cite{levenberg1944, marquardt1963}. This algorithm combines the Gauss-Newton method and gradient descent to optimize the curve fit iteratively. The LMA continues iterating until the relative change in the reduced chi-squared value or the difference in the control parameter between two successive iterations is less than a threshold of $10^{-8}$. To evaluate the goodness of fit, the coefficient of determination $R^2$ is calculated, where $R^2=1$ indicates a perfect fit.

\bibliography{s1sources}

\end{document}